\documentstyle[12pt]{article}
\begin{document}
\begin{flushright}
RAL-TR/95-048
\end{flushright}

\begin{center}
{\LARGE {\bf Unitarity and the Time Evolution of}}\\[0.3cm]
{\LARGE {\bf  Quantum Mechanical States }}\\[2.cm]
{\large P.K.~Kabir}\footnote[1]{E-mail address:
pkk@physics.virginia.edu}{\large ~and A.~Pilaftsis}\footnote[2]{E-mail
address: pilaftsis@v2.rl.ac.uk}\\[0.4cm]
{\it Rutherford Appleton Laboratory, Chilton, Didcot, Oxon, OX11 0QX,
UK}\\[0.3cm]
{\it and}\\[0.3cm]
{\it Institute of Nuclear and Particle Physics, J.W.~Beams
Laboratory of Physics,\\
University of Virginia, Charlotessville, VA 22901, USA}
\\[0.3cm]
\end{center}
\vskip2cm
\centerline{\bf ABSTRACT}
The basic requirement that, in quantum theory, the time-evolution of
any state is determined by the action of a unitary operator, is shown
to be the underlying cause for certain ``exact" results which
have recently been reported about the time-dependence of transition
rates in quantum theory. Departures from exponential decay, including
the ``Quantum Zeno Effect", as well as a theorem by Khalfin about
the ratio of reciprocal transition-rates, are shown to follow directly
from such considerations. At sufficiently short times, unitarity requires
that reciprocity must hold, independent of whether $T$-invariance is valid.
If  $T$-invariance does not hold, unitarity restricts the form of possible
time-dependence of reciprocity ratios.
\newpage

\section*{I. Introduction}

\indent
The Weisskopf-Wigner theory~\cite{WW} of decaying states has been used with
great success in a wide variety of applications. Nevertheless, since
it is an approximate theory, it is not surprising that there should be
circumstances in which one expects~\cite{LK,JS,CSM} departures from the
predictions of the theory. Some of these issues have acquired renewed interest
because of advances in experimental methods~\cite{IHBW}; others arise from the
expected~\cite{VLF,PKK}
deviation from time-reversal symmetry in weak interactions.
The question of the correct treatment of unstable particles also arises in the
application of current gauge theories of weak interactions, where the
instability
of intermediate bosons and fermions cannot always be neglected~\cite{AP}.
In this note, we show that many of these corrections can be directly
traced back to the fundamental requirement of unitarity, which is satisfied
only approximately in the Weisskopf-Wigner method.

In the Weisskopf-Wigner approximation ---which can be generalized~\cite{BL} to
the case of decays arising from two or more states---  certain initial
states are singled out for special attention. Transitions from these
distinguished states, to other states, deplete the population of these
initial states. The Weisskopf-Wigner approximation allows for this by
replacing the matrix-elements of the exact (full) Hamiltonian, in the
subspace spanned by those states, by a non-Hermitian submatrix.
For a single unstable state, the (negative) imaginary
part of the complex ``energy" assures that the probability decreases
exponentially with time. While
this prescription accounts for the ``leakage" of probability in terms
of the rate of transitions out of the initial states, detailed analysis
outlined below, shows that the Weisskopf-Wigner procedure cannot satisfy
unitarity exactly. The circumstance~\cite{LK} that the ``law"~\cite{RS}
of exponential decay {\em cannot} be exactly right in quantum theory can be
directly related to the fact that the exponential Ansatz is incompatible with
the unitarity requirement which
is essential for the basic interpretation of the theory. In this note, the
general condition imposed on transition amplitudes by the restriction of
unitarity is explicitly stated. When applied to a theorem about tests of
reciprocity originally given by Khalfin~\cite{LKCS}, one obtains not only a
simpler and more direct proof of the theorem but also a stipulation on
the nature of the variation whose occurrence Khalfin could infer, but
not specify further. The present formulation of the unitarity conditions
could be used to explicitly take account of this constraint in possible
future attempts to improve on the Weisskopf-Wigner approximation.

Section II presents the unitarity conditions which the exact
transition amplitudes must satisfy, and shows how these lead to
useful results, in addition to providing a new and simpler proof
of Khalfin's theorem. Section III summarizes our conclusions.

\section*{II. Unitarity Constraint on Transition Amplitudes}

{\em Theorem.} If $A_{kj}(t)$ is the exact transition amplitude for a state
initially prepared in the state $j$ to be found in the state $k$ after
a lapse of time $t$, unitarity requires that
\begin{equation}
                    A^*_{jk}(-t)\ =\  A_{kj} (t).                 \label{1}
\end{equation}
Correspondingly, if $f_{kj}(t) = A_{kj}(t)/A_{jk}(t)$, the function
$f_{kj}(t)$ should satisfy the relation
\begin{equation}
                 f^*_{kj}(-t)f_{kj}(t)\ =\ 1.          \label{2}
\end{equation}
{\em Proof.} By general principles of quantum mechanics, (using units
$\hbar=1$)
\begin{equation}
    A_{kj}(t)\ =\ \langle k| \exp (-iHt) |j\rangle             \label{3}
\end{equation}
for any two states $k$ and $j$, where $H$ is the complete
Hamiltonian governing the time-evolution of the system.

Hermiticity of $H$ assures that the operator $U(t) \equiv \exp (-iHt)$
is unitary since $[\exp (-iHt)]^\dagger=\exp(iHt)$. Thus
\begin{displaymath}
       \langle k| U |j \rangle  \equiv \langle k | \exp (-iHt) |j\rangle\ =\
    \langle j| U^\dagger|k\rangle ^*\ =\  \langle j| \exp(+iHt) |k\rangle^*.
\end{displaymath}
{}From Eq.\ (\ref{3}), the conjugated quantity on the R.H.S. is just
$A_{jk}(-t)$.
Consequently,
\begin{displaymath}
                    A_{kj}(t)\ =\  A^*_{jk} (-t)
\end{displaymath}
for transitions induced by any Hermitian Hamiltonian $H$, which is
exactly Eq.~(\ref{1}).   This may be regarded
as the unitarity constraint on transition amplitudes. If we rewrite
Eq.~(\ref{1}) in the form:
\begin{equation}
    \phi_{jk} \equiv   \frac{A^*_{jk}(-t)}{ A_{kj}(t)}\ =\ 1,     \label{4}
\end{equation}
which must be valid for any $j,k$, then the condition
$\phi_{jk}=\phi_{kj}$ leads to
\begin{displaymath}
               \frac{A^*_{jk}(-t)}{ A_{kj}(t)}\ =\               \label{4a}
                \frac{A^*_{kj}(-t)}{A_{jk}(t)}   ,
\end{displaymath}
which is equivalent to Eq.~(\ref{2}). \qquad\qquad\qquad
\qquad\qquad\qquad\qquad\qquad\qquad\qquad    {\em Q.E.D.}

Our first application of the theorem will be to use it to show
that the decay probability of an unstable state must be an $even$
function of time. Setting $j=k$ in Eq.~(\ref{1}), we obtain
\begin{equation}
P_{jj}(-t)\ =\ | A_{jj}(-t) |^2\ =\ | A^*_{jj}(t) |^2\ =\ P_{jj}(t).
   \label{5}
\end{equation}
Therefore, the probability for a quantum system to remain in its
initial state, and consequently also the complementary probability
to make transitions to other states, must be an even function of
time. This result has been known for some time, even though it has
not yet found its way into many textbooks. The symmetry of $P_{jj}$
under $ t \to -t $ is even more apparent if one writes
\begin{equation}
 P_{jj} (t)\ =\ A^*_{jj}(t)\, A_{jj}(t)\ =\ A_{jj}(-t)\, A_{jj}(t), \label{5'}
\end{equation}
making use of Eq.~(\ref{1}). Provided that $P_{jj}(t)$ is differentiable ---a
condition which is assured if
$\langle H\rangle $ exists for the given initial state---
at $t=0$, it follows that $\dot P_{jj}(t)$, which must correspondingly be an
odd function of time, must vanish at $t=0$. We can explicitly verify this
by calculating
\begin{equation}
    P_{kj}(t)\ =\ | A_{kj}(t) |^2\ =\ A^*_{kj}(t)\, A_{kj}(t)\  =\
A_{jk}(-t)\, A_{kj}(t).
   \label{6}
\end{equation}
Then
\begin{equation}
    \dot P_{kj}(t)\ =\ \dot A_{jk}(-t)A_{kj}(t) + A_{jk}(-t) \dot A_{kj}(t),
\end{equation}
and, if we substitute the explicit expressions from Eq.~(\ref{3}), we obtain
\begin{equation}
    \dot P_{kj}(t)\ =\ i \langle j| He^{iHt}|k\rangle A_{kj}(t) -
                     iA^*_{kj}(t)\langle k|He^{-iHt}|j \rangle
\end{equation}
and thus
\begin{equation}
\dot P_{kj}(0)\ =\ 2 \Im m \Big( \langle j|k \rangle\langle k|H|j\rangle \Big),
\label{10}
\end{equation}
which yields~\cite{f2}
\begin{equation}
   \sum_{k} \dot P_{kj}(0)\ =\ 2 \Im m \langle j|H|j\rangle\ =\ 0. \label{5''}
\end{equation}
This constraint has been called~\cite{MS} the Quantum Zeno Effect.
Observation~\cite{IHBW} of the expected non-linear time-dependence at short
times, for the closely related process of $induced$ transitions, can be
regarded as evidence in support of the effect.

A corollary statement is that, since $\exp (-\gamma t)$ does $not$ have a
vanishing derivative at $t=0$, the hypothesis of exponentially decaying states
is {\em inconsistent} with the requirement of unitarity. The deviations
from exponential decay, both at very short times and at very long
times, have been extensively studied by many authors~\cite{LK,CSM}.

Our next use of the theorem will be to provide a new and simpler proof of
Khalfin's theorem~\cite{LKCS}:
that if the ratio of the transition amplitudes for two
reciprocal~\cite{ff}
transitions $a \to b$ and $b \to a$ is constant, then the only
possible value for the modulus $R$ of that constant is unity.
We have seen that $f_{kj}(t)$, defined after Eq.~(\ref{1}), must satisfy
\begin{displaymath}
                  f_{kj}(t)f^*_{kj}(-t)\ =\ 1 .
\end{displaymath}
Thus, if we are given that $f_{kj}(t)=\zeta =${\em constant} for all $t\ge 0$,
it follows from above that $f^*_{kj}$ must also be constant for all
$t \le 0 $.
Continuity of $|f_{kj}|$ at  $t=0$ requires that
\begin{equation}
           R\ \equiv\   |\zeta|\ =\ 1 ,                          \label{7}
\end{equation}
which is Khalfin's theorem.
Since Khalfin arrived at this conclusion by a more complicated argument, we
should like to note that the proof presented here required little~\cite{FF}
more
than the assumption of unitarity. In particular, no assumption is required
about the positivity of the spectrum of $H$, viz.\ the assumption Spec$\, H
\ge 0$, made in Khalfin's proof, appears to be unnecessary.

Our next application of these ideas will be to prove that reciprocity
must hold (independent of the question of time-reversal invariance)
at very short times, as a consequence of unitarity alone.   From
Eq.~(\ref{1}),
\begin{equation}
             P_{jk}(0)\ =\ | A_{jk}(0) |^2\ =\ | A^*_{kj}(0)|^2\ =\ P_{kj}(0),
    \label{6b}
\end{equation}
which states that a kind of reciprocity is exactly valid at $t = 0$. This can
be understood directly as follows. For small values of $t$, let us expand the
RHS of Eq.~(\ref{3}) in a power series,
\begin{equation}
          A_{kj}(t)\ =\ \langle k| 1\ -\ iHt\ +\ \cdots |j\rangle .  \label{7a}
\end{equation}
If the state $k$ is not orthogonal to $j$, the value of the RHS of
Eq.~(\ref{7a}),
for $t=0$, is the complex conjugate of, and therefore has the same
magnitude as, $\langle j|k\rangle $. Therefore, $P_{kj}(0)$ and $P_{jk}(0)$
must be
equal in that case. If $k$ is orthogonal~\cite{ff3} to $j$, we must go to the
term linear in $t$ in Eq.~(\ref{7a}), and we find that the transition
amplitude is proportional to $H_{kj}$. But, since $H$ must be Hermitian,
this is the complex conjugate of the matrix-element $H_{jk}$ for the
inverse transition and we recover the result, reported in many
textbooks~\cite{JJS}, that in lowest-order perturbation theory, reciprocity
follows from the Hermiticity of the interaction Hamiltonian.
This requirement of reciprocity, independent of the $T$-invariance or otherwise
of the Hamiltonian $H$, at early times $t \to 0$, can be stated more precisely
by expanding the $P_{kj}(t)$ as a power series in $t$:
\begin{equation}
P_{kj}(t)\ =\ P_{kj}(0)\ +\  \dot{P}_{kj}(0)\, t\ +\ \frac{1}{2}\,
\ddot{P}_{kj}(0)\, t^2\ +\ {\cal{O}} ( t^3 ) .
\end{equation}
We have already seen that $P_{kj}(0)= |\langle k|j\rangle |^2$ and
$ \dot P_{kj}(0)$, Eq.~(\ref{10}),
both vanish if $\langle k|j\rangle = 0$. By direct calculation, we find
\begin{equation}
 \ddot P_{kj}(0)\ =\ 2|\langle k |H|j\rangle |^2\ -\ 2\,
\Re e\Big[ \langle j|k \rangle \langle k|H^2|j\rangle\Big],
\end{equation}
which shows explicitly that reciprocity {\em must} be preserved if $\langle
k | j\rangle = 0 $, to order
$t^2$, solely as a consequence of the Hermiticity of $H$, viz.\
of the requirement of unitarity for the time-evolution operator.

By an application of this result to the argument which led to the ``quantum
Zeno's paradox", we can conclude that any departure from reciprocity, which
would be expected if $T$-invariance is not a symmetry of the underlying
Hamiltonian, will be reduced or suppressed if the system
undergoing change is monitored too closely. Thus, for example, frequent
observation, amounting to a measurement of its strangeness, of a neutral kaon
state, could reduce the  inferred value of the $CP$- and $T$-violating
parameter $\varepsilon$ (under the assumption of $TCP$-invariance) relative
to the one measured for ``free" kaons. Possible implications of the
corresponding quantum Zeno effect for baryogenesis in the Universe
will be discussed elsewhere.

 From his theorem Khalfin could conclude that, if reciprocity is {\em not}
satisfied, $R$ must vary with time although nothing further could be said
about the nature of that variation. The general solution for a function
satisfying the unitarity condition~(\ref{2}), can be written as:
\begin{equation}
                   f_{jk}(t)\ =\ \exp [g(t) + i h(t)]   ,      \label{8}
\end{equation}
where $g(t)$ and $h(t)$ are {\em real} functions of $t$ which must be
{\em odd} and {\em even}, respectively, under $t \to -t$.
Any phenomenological representation of $A_{jk}(t)$, and correspondingly of
$f_{jk}(t)$, to take account of possible deviations from reciprocity,
which conforms to Eq.~(\ref{8}), will automatically satisfy the
requirement of unitarity. The Weisskopf-Wigner formalism, as extended to the
case of interfering~\cite{BL} decaying states, was applied by Lee, Oehme, and
Yang~\cite{LOY} to the $K^0-\bar{K}^0$ system, ---in a form which can
accommodate
possible $T$-noninvariance---, and appears to adequately represent the data
obtained thus far. Notwithstanding its great success, our foregoing discussion
has shown that this description is not strictly compatible with unitarity of
the exact theory~\cite{fff4}.

Knowledge of the spectral content of the initial state, and
thereby of the spectrum of $H$, determines~\cite{KF}, in principle, the
complete time-evolution of the system through decomposition of its state-vector
into a
complete set of eigenvectors of $H$. Even in the absence of such detailed
knowledge, any additional information about the spectrum of $H$, which could
be expressed as further constraints~\cite{LAK} on the functions $g(t)$ and
$h(t)$ ---beyond the conditions on $g(t)$ mentioned already--- would obviously
help to define the admissible forms of time-dependence.

We have already seen above that the Weisskopf-Wigner exponential Ansatz
$cannot$ exactly satisfy unitarity. Eq.~(\ref{8}), with possible supplementary
conditions, offers a natural point of departure for a new phenomenology
satisfying exact unitarity.

\section*{III. Conclusions}

            In this note, we have shown that the existence of certain puzzling
and unexpected phenomena, such as the quantum Zeno effect or deviations from
Rutherford's law of exponential decay, can be directly traced back to the
{\em unitarity} condition, which is required in quantum theory for a consistent
description of any (isolated) dynamical system. Further consequences which are
expected in principle, in addition to the Khalfin theorem mentioned already,
include an analogue of the Zeno effect for the comparison of rates of
reciprocal
transitions. Whereas these rates are {\em not} directly related unless
$T$-invariance is
imposed on {\em all} relevant interactions, unitarity alone requires that a
test of reciprocity {\em must} yield a result conforming to the $T$-invariant
expectation {\em if}
the measurements are made sufficiently rapidly. This means, for example, that
even if we accept the usual interpretation~\cite{VLF} that the observed
$CP$-noninvariance observed in neutral K-meson decays is associated with a
$T$-noninvariant interaction, the corresponding expected~\cite{PKK} {\em
departure}
from reciprocity in $K^{0} \rightleftharpoons \bar K^{0}$ transitions would be
suppressed, and indeed disappear, if the comparison were made at shorter and
shorter times. Such asymmetries have been invoked~\cite{ADS} to explain the
observed
baryon asymmetry of the Universe; possible implications of this ``$CP$
and $T$ quantum Zeno effect'' will be discussed elsewhere.

\section*{Acknowledgments}
One of us (PK) thanks H.M. Chan and F.E. Close for kind hospitality at the
Rutherford-Appleton Laboratory, and X. Tata for helpful comments. This research
is supported in part by the U.S.\ Department of Energy.


\begin{thebibliography}{99}

\bibitem{WW} V.\ Weisskopf and E.\ Wigner, {\em Z.~Phys.} {\bf 63}, 54 (1930);
{\em ibid.}~{\bf 65}, 18 (1930);
      for a compact account, see W.~Heitler, Quantum Theory of Radiation,
      Oxford U.P., 3rd.~ed.,1954.

\bibitem{LK} L.\ Khalfin, {\em Zh.~Eksper.~Teor.~Fiz.}~{\bf 33}, 1371 (1957)
[{\em Sov.~Phys.~JETP}~{\bf 6}, 1053 (1958)].

\bibitem{JS}  {\em E.g.}, J. Schwinger, {\em Ann. Phys.}~{\bf 9}, 169 (1958).

\bibitem{CSM} C.B.~Chiu, E.C.G.~Sudarshan, and B.~Misra,
{\em Phys.\ Rev.}\ {\bf D16}, 520 (1977);
C.\ Bernardini, L.\ Maiani, and M.\ Testa, {\em Phys.\ Rev.\ Lett.}\ {\bf 71}
2687 (1993);
G.\ Garc\'ia-Calder\'on, J.L.\ Mateos, and M.\ Moshinsky,
{\em Phys.\ Rev.\ Lett.}\ {\bf 74}, 337 (1995).

\bibitem{IHBW} W.M.~Itano, D.J.~Heinzen, J.J.~Bollinger,
and D.J.~Wineland, {\em Phys.\ Rev.}\ {\bf A41}, 2295 (1990).

\bibitem{VLF}  V.L.\ Fitch, {\em Rev.\ Mod.\ Phys.}\ {\bf 53}, 367 (1981).

\bibitem{PKK} P.K.\ Kabir, {\em Phys. Rev.\ }{\bf D2}, 540 (1970).

\bibitem{AP} A.\ Pilaftsis, {\em Z.\ Phys.}\ {\bf C47}, 95 (1990);
J.\ Papavassiliou and A.\ Pilaftsis, {\em Gauge Invariance and Unstable
Particles}, Rutherford report (1995), RAL-TR-95-021, {\em Phys.\ Rev.\ Lett.}\
(to appear).


\bibitem{BL} G.\ Breit and I.S.\ Lowen, {\em Phys. Rev.}\ {\bf 46}, 590 (1934).

\bibitem{RS} E.\ Rutherford and F.~Soddy, {\em Phil.~Mag.}~{\bf 6}, 576 (1903).

\bibitem{LKCS} L.~Khalfin, unpublished Univ.~of
         Texas Report DOE-ER 40200-211, 1990,
         cited by C.B.~Chiu and E.C.G.~Sudarshan, {\em Phys.\ Rev.}\
{\bf D42}, 3712 (1990).

\bibitem{f2} Strictly speaking, the initial state $j$ should be excluded from
the sum over final states $k$. The RHS of Eq.~(\ref{5''}) is not affected by
this restriction. This confirms the result of the direct
calculation of $\dot{P}_{jj}(0)$ in Ref.~\cite{MS}.

\bibitem{MS} B.\ Misra and E.C.G.\ Sudarshan, {\em J.\ Math.\ Phys.}\ {\bf 18},
756 (1977).

\bibitem{ff} For the present, following Khalfin, we restrict our discussion to
states $a,\ b$ which can be transformed into the corresponding
``time-reversed"
states $a_{T},\ b_{T}$ by rotation and/or other symmetry of the governing
Hamiltonian  $H$.

\bibitem{FF} To assure the existence of the transition rates, it would
suffice,
for example, if the initial and final states had finite expectation values of
$\langle H\rangle $.

\bibitem{ff3} A convenient way to think of this is to imagine that $k$ and
$j$ are distinct stationary states of some Hamiltonian $H_{0}$ and that
transitions between them are induced by $H' = H - H_{0}$.

\bibitem{JJS}
{\em E.g.}, J.J.\ Sakurai, Modern Quantum Mechanics, Addison-Wesley,
 New York, 1992.

\bibitem{LOY} T.D.~Lee, R.~Oehme, and C.N.~Yang, {\em Phys. Rev.}\ {\bf 106},
 340  (1957).

\bibitem{fff4} Bell's proof~\cite{JB}, that unitarity can be satisfied in
a generalized Weisskopf-Wigner theory, applies to a system whose time-evolution
is exactly determined by a non-Hermitian (time-independent) {\em effective}
Hamiltonian, viz.\ whose ``eigenstates'' decay exponentially. Since
we have seen that truly exponential decay cannot occur for any admissible
(Hermitian) Hamiltonian, Bell's demonstration does not carry over to the
exact (complete) theory.

\bibitem{JB} J.S.\ Bell, in ``High Energy Physics'', C.\ de Witt and
M.\ Jacob, eds., Gordon $\&$ Breach, New York, 1965.


\bibitem{KF} N.S.\ Krylov and V.A.\ Fock, {\em Zh.\ Eksp.\ Teor.\ Fiz.}\
{\bf 17}, 93 (1947).

\bibitem{LAK} Apart from the requirements of unitarity, further
restrictions may follow from other physical requirements, L.A.\ Khalfin,
private communication.


\bibitem{ADS} A.D.\ Sakharov, {\em Pis'ma Zh.\ Eksp.\ Teor.\ Fiz.}\ {\bf 5}, 32
(1967),\ [{\em JETP Letters}\ {5}, 24 (1967)].



\end{thebibliography}
\end{document}